\documentclass[amsmath,amssymb,aps,prd,showpacs,showkeys,nofootinbib]{revtex4}
\usepackage{graphicx}
\usepackage{dcolumn}
\usepackage{bm}
\usepackage{latexsym,color}
\usepackage{natbib}
\usepackage{hyperref}
\usepackage{pstricks}

\def\mnras{Mon. Not. Roy. Astr. Soc.}
\def\apj{Astrophys. J.}
\def\apjl{Astrophys. J. Lett.}

\def\aap{Astron. Astrophys.}

\def\prd{Phys. Rev. D}

\begin{document}
\def\be{\begin{equation}}
\def\ee{\end{equation}}
\def\bea{\begin{eqnarray}}
\def\eea{\end{eqnarray}}
\title{Gravitational field of compact objects in general relativity}

\author{Kuantay Boshkayev$^{1,2}$, Hernando Quevedo$^{1,3}$ and Remo Ruffini$^{1}$}
\affiliation{
$^1$Dipartimento di Fisica and ICRA, Universit\`a di Roma "La Sapienza´´,  Piazzale Aldo Moro 5, I-00185 Roma, Italy\\
$^2$Department of Physics, Al-Farabi Kazakh National University, Al-Farabi avenue 050038, Almaty, Kazakhstan\\
$^3$Instituto de Ciencias Nucleares, Universidad Nacional Aut\'onoma de M\'exico, AP 70543, M\'exico, DF 04510, Mexico}
\email{kuantay@mail.ru, quevedo@nucleares.unam.mx, ruffini@icra.it}

\date{\today}

\begin{abstract}
We study some exact and approximate solutions of Einstein's equations that can be used to describe the gravitational field of astrophysical compact objects in the limiting case of slow rotation and slight deformation. First, we show that none of the standard models obtained by using Fock's method can be used as an interior source for the approximate exterior Kerr solution. We then use Fock's method to derive a generalized interior solution, and also an exterior solution that turns out to be equivalent to the exterior Hartle-Thorne approximate solution that, in turn, is equivalent to an approximate limiting case of the exact Quevedo-Mashhoon solution. As a result we obtain an analytic approximate solution that describes the interior and exterior gravitational field of a slowly rotating and slightly deformed astrophysical object.
\end{abstract}


\maketitle

\section{Introduction}
\label{intro}

In astrophysics, the term compact object is used to refer to objects which are small for their mass. In a wider sense, the class of compact objects is often defined to contain collectively planet-like objects, white dwarfs, neutron stars, other exotic dense stars, and black holes. It is well known that Newtonian theory of gravitation provides an adequate description of the gravitational field of conventional astrophysical objects. However, the discovery of exotic compact objects such as quasars and pulsars together with the possibility of continued gravitational collapse to a black hole points to the importance of relativistic gravitation in astrophysics. Moreover, advances in space exploration and the development of modern measuring techniques have made it necessary to take relativistic effects into account even in the Solar system. Probably the simplest way to study the relativistic gravitational field of astrophysical compact objects is by expressing it in terms of their multipole moments, in close analogy with the Newtonian theory, taking into account the rotation and the internal structure of the source.

In this context, the first exterior solution with only a monopole moment was discovered by Schwarzschild \cite{1916AbhKP......189S}, soon after the formulation of Einstein's theory of gravity. In 1917, Weyl \cite{1917AnP...359..117W} showed that the problem of finding static axisymmetric vacuum solutions can generically be reduced to a single linear differential equation whose general solution can be represented as an infinite series. The explicit form of this solution resembles the corresponding solution in Newtonian's gravity, indicating the possibility of describing the gravitational field by means of multipole moments. In 1918, Lense and Thirring \cite{1918PhyZ...19..156L} discovered an approximate exterior solution which, apart from the mass monopole, contains an additional parameter that can be interpreted as representing the angular momentum of the massive body. From this solution it became clear that, in Einstein's relativistic theory, rotation generates a gravitational field that leads to the dragging of inertial frames (Lense-Thirring effect). This is the so--called gravitomagnetic field which is of especial importance in the case of rapidly rotating compact objects. The case of a static axisymmetric solution with monopole and quadrupole moment was analyzed in 1959 by Erez and Rosen \cite{erez1959} by using spheroidal coordinates which are specially adapted to describe the gravitational field of non-spherically symmetric bodies. The exact exterior solution which considers arbitrary values for the angular momentum was found by Kerr \cite{Kerr1963PhRvL..11..237K} only in 1963. The problem of finding exact solutions changed dramatically after Ernst \cite{1968PhRv..167.1175E} discovered in 1968 a new representation of the field equations for stationary axisymmetric vacuum solutions. In fact, this new representation was the starting point to investigate the Lie symmetries of the field equations. Today, it is known that for this special case the field equations are completely integrable and solutions can be obtained by using the modern solution generating techniques \cite{1982RSPSA.382..221D}.
A comprehensive review on solution generating techniques and stationary axisymmetric global solutions of Einstein and Einstein-Maxwell equations is given in \cite{1985cup..book.....I}. There are several solutions with higher multipole moments \cite{1990CQGra...7..779C, 1990CQGra...7L.209M, 1992CQGra...9.2477M, 2000PhRvD..61h1501M,2006PhRvD..73j4038P} with very interesting physical properties. In this work, we will analyze a particular class of solutions, derived by Quevedo and Mashhoon \cite{1991PhRvD..43.3902Q} in 1991, which in the most general case contains infinite sets of gravitational and electromagnetic multipole moments. Hereafter this solution will be denoted as the QM solution.

As for the interior gravitational field of compact objects, the situation is more complicated. There exists in the literature a reasonable number of interior spherically symmetric solutions \cite{1960PhRv..120..321A} that can be matched with the exterior Schwarzschild metric. Nevertheless, a major problem of classical general relativity consists in finding a physically reasonable interior solution for the exterior Kerr metric. Although it is possible to match numerically the Kerr solution with the interior field of an infinitely tiny rotating disk of dust \cite{1993ApJ...414L..97N}, such a solution cannot be used in general  to describe astrophysical compact objects. It is now widely believed that the Kerr solution is not appropriate to describe the exterior field of rapidly rotating compact objects. Indeed, the Kerr metric takes into account the total mass and the angular momentum of the body. However, the quadrupole moment is an additional characteristic of any realistic body which should be considered in order to correctly describe the gravitational field. As a consequence, the multipole moments of the field created by a rapidly rotating compact object are different from the multipole moments of the Kerr metric \cite{stergioulas}. For this reason a solution with arbitrary sets of multipole moments, such as the QM solution, can be used to describe the exterior field of arbitrarily rotating mass distributions.

In the case of slowly rotating compact objects it is possible to find approximate interior solutions with physically meaningful energy-momentum tensors and state equations. Because of its physical importance, in this work we will review the Hartle-Thorne \cite{H1967,HT1968} interior solution which are coupled to an approximate exterior metric. Hereafter this solution will be denoted as the HT solution. One of the most important characteristics of this family of solutions is that the corresponding equation of state has been constructed using realistic models for the internal structure of relativistic stars. Semi-analytical and numerical generalizations of the HT metrics with more sophisticated equations of state have been proposed by different authors \cite{2001CQGra..18..969A}. A comprehensive review of these solutions is given in \cite{stergioulas}. In all these cases, however, it is assumed that the multipole moments (quadrupole and octupole) are relatively small and that the rotation is slow.

To study the physical properties of solutions of Einstein's equations, Fock \cite{fock1959} proposed an alternative method in which the parameters entering the exterior metric are derived by using physical models for the internal structure of the body. In this manner, the significance of the exterior parameters become more plausible and the possibility appear of determining certain aspects of the interior structure of the object by using observations performed in the exterior region of the body. Fock's metric in its first-order approximation was recently generalized in 1985 by Abdildin \cite{1985pft..conf...20A, abdildin2006} (for details see Appendix \ref{sec:appabd}).

In this work, we review the main exact and approximate metrics which can be used to study the interior and exterior gravitational field of compact objects and find the relationships between them. We will show that the exterior HT approximate solution is equivalent to a special case of the QM solution in the limit of a slowly rotating slightly deformed compact object (first order in the quadrupole and second order in the angular momentum). Moreover, we will show that a particular case of the extended Fock metric is equivalent to the approximate exterior HT solution.
Furthermore, since those particular cases of the exterior HT metric that possess internal counterparts with plausible equations of state are also special cases of the exterior QM metric, we conclude that at least in those particular cases it should be possible to match the QM solution with an exact interior still unknown solution so that it describes globally the gravitational field of astrophysical compact bodies.

This paper is organized as follows. In Section  \ref{sec:ht} we review the HT  solutions and briefly comment on their most important properties. In Section  \ref{sec:fock} we present Fock's extended metric, as first derived by Abdildin \cite{1985pft..conf...20A, abdildin2006} in harmonic coordinates, and introduce a set of new coordinates which makes it suitable for comparison with other exterior metrics. Moreover, we find explicitly the coordinate transformation that  relates Fock's extended metric with the exterior HT  solution.

In Section  \ref{sec:qm} we present a particular case of the QM metric which contains, in addition to the mass and angular momentum parameters, an additional parameter related to the mass quadrupole of the source. Here we show explicitly that a limiting case of the QM metric contains the HT metric. Finally, Section \ref{sec:con} contains discussions of our results and suggestions for further research.

\section{The Hartle-Thorne metrics}
\label{sec:ht}

To second order in the angular velocity, the structure of compact objects can be approximately described by the mass, angular momentum and quadrupole moment. An important consequence of this approximation is that the equilibrium equations reduce to a set of ordinary differential equations. Hartle and Thorne \cite{H1967, HT1968} explored the gravitational field of rotating stars in this slow rotation approximation. This formalism can be applied to most compact objects including pulsars with millisecond rotational periods, but it shows \lq\lq large" discrepancies in the case of rapidly rotating relativistic objects near the mass-shedding limit according to \cite{stergioulas}, i. e., when the angular velocity of the object reaches the angular velocity of a particle in a circular Keplerian orbit at the equator.
In fact, recently in \cite{berti2004} and \cite{berti2005} it was shown that the second order rotation corrections of the HT metric are sufficient to describe the properties of stars with intermediate rotation rates.  These results were generalized in \cite{benhar2005} to include third order corrections. It turns out that third order corrections are irrelevant at the mass-shedding limit; however, they are important to study the moment of inertia of rapidly rotating neutron stars. Moreover, in \cite{2006PhRvD..73j4038P} an analytical solution was derived that can be matched accurately with interior numerical solutions. On the other hand, an alternative numerical study \cite{2011ApJ...728...12L} shows that in the case of uniformly rotating neutron stars the dimensionless specific angular momentum cannot exceed the value ~0.7.

An additional property of this formalism is that it can be used to match an interior solution with an approximate exterior solution. In this connection, it is worth noticing that the problem of matching interior and exterior solutions implies many mathematical and physical issues \cite{2008CQGra..25v8002P, 2009JPhCS.189a2028P, 2012PhRvL.108w1104P, 2012MNRAS.422.2581P, 2012ApJ...756...82P,2012mgm..conf...35Q}, including the performance of the metric functions and the coordinates at the matching surface as well as the physical behavior of the internal parameters like the density and pressure of the matter distribution.
In the following subsections we will present the interior and the exterior metrics and introduce notations which will be used throughout the paper.

\subsection{The interior solution}

If a compact object is rotating slowly, the calculation of its equilibrium properties reduces drastically because it can be considered as a linear perturbation of an already-known non-rotating configuration. This is the main idea of Hartle's formalism \cite{H1967}. To simplify the computation the following conditions are assumed to be satisfied.

\noindent
1) Equation of state: the matter in equilibrium configuration is assumed to satisfy a one-parameter equation of state, $\mathcal{P}=\mathcal{P}(\mathcal{E})$, where $\mathcal{P}$ is the pressure and $\mathcal{E}$ is the mass-energy density.

\noindent
2) Axial and reflection symmetry:   the configuration is symmetric with respect to an arbitrary axis which can be taken as the rotation axis. Furthermore, the rotating object should be invariant with respect to reflections about a plane perpendicular to the axis of rotation.

\noindent
3) Uniform rotation:   only uniformly rotating configurations are considered since it is known that configurations that  minimize the total mass-energy (e.g., all stable configurations) must rotate uniformly \cite{HS1967} \footnote{ Notice, however, that stability is also possible in the case differentially rotating configurations. See, for instance, \cite{1989MNRAS.237..355K, 1989MNRAS.239..153K}}.

\noindent
4) Slow rotation:  this means that angular velocities $\Omega$ are small enough so that the fractional changes in pressure, energy density and gravitational field due to the rotation are all less than unity, i.e.
\begin{equation}
\Omega^2 \ll \left(\frac{c}{R'}\right)^2 \frac{G M'}{c^2 { R'}}
\end{equation}
where $ M'$ is the mass and $R'$ is the radius of the non-rotating configuration. The above condition is equivalent to the physical requirement $\Omega\ll c/ R'$.

When the equilibrium configuration described above is set into slow rotation, the geometry of space-time around it and its interior distribution of stress-energy are changed. With an appropriate choice of coordinates, the perturbed geometry is described by
\begin{equation}\label{h1}
\begin{split}
ds^{2}=e^{\nu}\left[1+2(h_0+h_2P_2)\right]dt^{2}-\frac{\left[1+2(m_0+m_2P_2)/(R-2M')\right]}{1-2 M'/R}dR^2-
\\-R^2\left[1+2(v_2-h_2)P_2\right]\left[d\Theta^2+\sin^2\Theta(d\phi-\omega dt)^2\right]+O(\Omega^3)
\end{split}
\end{equation}

Here $M'$ is the mass of the non-rotating star, $P_2=P_2(\cos\Theta)$ is the Legendre polynomial of second order, $\omega$ is the angular velocity of the local inertial frame, which is a function of $R$ and is proportional to the star's angular velocity $\Omega$, and, finally, $h_0, h_2, m_0, m_2, v_2$ are all functions of $R$ that are proportional to $\Omega^2$.

In the above coordinate system the fluid inside the star moves with a 4-velocity corresponding to a uniform and rigid rotation \cite{Thorne1967}. The  contravariant components are
\begin{equation}\label{h2}
u^t=(g_{tt}+2\Omega g_{t\phi}+\Omega^2 g_{\phi\phi})^{-1/2}, \quad u^\phi=\Omega u^t, \quad u^R=u^\Theta=0.
\end{equation}
The quantity $\Omega$ that appears in the expression for $u^t$ is so defined that
$\bar{\omega}\equiv\Omega-\omega$ is the angular velocity of the fluid relative to the local inertial frame.

The energy density $\mathcal E$ and the pressure $\mathcal P$ of the fluid are affected by the rotation because it deforms the compact object. In the interior of the object at a given $(R,\Theta)$, in a reference frame that is momentarily moving with the fluid, the pressure and the density of mass-energy are
\begin{eqnarray}\label{h4}
\mathcal P&\equiv& P+(E+P)(p_0^*+p_2^*P_2)= P+\Delta P\,\\
\mathcal E&\equiv& E+(E+P)(dE/dP)(p_0^*+p_2^*P_2)= E+\Delta E.
\end{eqnarray}
Here, $p_0^*$ and $p_2^*$ are dimensionless functions of $R$ that are proportional to $\Omega^2$, and describe the pressure perturbation, $P$ is the pressure and $E$ is the energy density of the non-rotating configuration.  The stress-energy tensor for the fluid of the rotating object is
\begin{equation}\label{h6}
T_\mu^{\nu}=(\mathcal E+\mathcal P)u_{\mu}u^{\nu}-\mathcal P\delta_{\mu}^{\nu}.
\end{equation}
The rotational perturbations of the objects's structure are described by the functions $\bar{\omega}, h_0, m_0, p_0^*, h_2, m_2, v_2, p_2^*$. These functions are calculated from Einstein's field equations (for details see \cite{H1967, HT1968}).

\subsection{The Exterior Solution}
\label{sec:htext}

The HT metric describing the exterior field of a slowly rotating slightly deformed object is given by
\begin{equation}\label{ht1}
\begin{split}
ds^2=\left(1-\frac{2{\mathcal M }}{R}\right)\left[1+2k_1P_2(\cos\Theta)+2\left(1-\frac{2{\mathcal M}}{R}\right)^{-1}\frac{J^{2}}{R^{4}}(2\cos^2\Theta-1)\right]dt^2
\\-\left(1-\frac{2{\mathcal M}}{R}\right)^{-1}\left[1-2k_2P_2(\cos\Theta)-2\left(1-\frac{2{\mathcal M}}{R}\right)^{-1}\frac{J^{2}}{R^4}\right]dR^2
\\-R^2[1-2k_3P_2(\cos\Theta)](d\Theta^2+\sin^2\Theta d\phi^2)+\frac{4J}{R}\sin^2\Theta dt d\phi\,
\end{split}
\end{equation}
with
\begin{eqnarray}\label{ht2}
k_1&=&\frac{J^{2}}{{\mathcal M}R^3}\left(1+\frac{{\mathcal M}}{R}\right)+\frac{5}{8}\frac{Q-J^{2}/{\mathcal M}}{{\mathcal M}^3}Q_2^2\left(\frac{R}{{\mathcal M}}-1\right)\ , \quad k_2=k_1-\frac{6J^{2}}{R^4}\ , \nonumber\\
k_3&=&k_1+\frac{J^{2}}{R^4}+\frac{5}{4}\frac{Q-J^{2}/{\mathcal M}}{{\mathcal M}^2R}\left(1-\frac{2{\mathcal M}}{R}\right)^{-1/2}Q_2^1\left(\frac{R}{\mathcal M}-1\right)\ ,\nonumber
\end{eqnarray}
where
\begin{equation}\label{legfunc}
Q_{2}^{1}(x)=(x^{2}-1)^{1/2}\left[\frac{3x}{2}\ln\frac{x+1}{x-1}-\frac{3x^{2}-2}{x^{2}-1}\right],
\ \ Q_{2}^{2}(x)=(x^{2}-1)\left[\frac{3}{2}\ln\frac{x+1}{x-1}-\frac{3x^{3}-5x}{(x^{2}-1)^2}\right],
\end{equation}
are the associated Legendre functions of the second kind. The constants ${\mathcal M}$, $J$ and $Q$ are related to the total mass, angular momentum and mass quadrupole moment of the rotating object,  respectively. This form of the metric corrects some misprints of the original paper by Hartle and Thorne \cite{HT1968} (see also \cite{bini2009} and \cite{berti2005}).

The total mass of a rotating configuration is defined as $ \mathcal M=M'+\delta M$, where $M'$ is the mass of non-rotating configuration and $\delta M$ is the change in mass of the rotating from the non-rotating configuration with the same central density. It should be stressed that in the terms involving $J^2$ and $Q$ the total mass $\mathcal M$ can be substituted by $M'$ since $\delta M$ is already a second order term in the angular velocity.

In general, the HT metric represents an approximate vacuum solution, accurate to second order in the angular momentum $J$ and to first order in the quadrupole parameter $Q$. In the case of ordinary stars, such as the Sun, considering the gravitational constant $G$ and the speed of light $c$, the metric (\ref{ht1}) can be further simplified due to the smallness of the parameters: \be\label{ht3} \frac{G\mathcal M_{Sun}}{c^{2}\mathcal R_{Sun}}\approx 2\times 10^{-6},\qquad \frac{G J_{Sun}}{c^{3}\mathcal R_{Sun}^{2}}\approx 10^{-12},\qquad \frac{G Q_{Sun}}{c^{2}\mathcal R_{Sun}^{3}}\approx 10^{-10}. \ee

For this special case one can calculate the corresponding approximate metric from (\ref{ht1}) in the limit $c\rightarrow\infty$.  The computations are straightforward and lead to
\begin{equation}
\label{ht4}
\begin{split}
ds^2=\left[1-\frac{2G{\mathcal M}}{c^{2}R}+\frac{2GQ}{c^{2}R^{3}}P_2(\cos\Theta)+\frac{2G^{2}{\mathcal M }Q}{c^{4}R^4}P_2(\cos\Theta)\right]c^{2}dt^2 +\frac{4GJ}{c^{2}R}\sin^2\Theta dt d\phi \qquad\qquad
\\-\left[1+\frac{2G{\mathcal M }}{c^{2}R}-\frac{2GQ}{c^{2}R^{3}}P_2(\cos\Theta)\right]dR^2
-\left[1-\frac{2GQ}{c^{2}R^{3}}P_2(\cos\Theta)\right]R^2(d\Theta^2+\sin^2\Theta d\phi^2)\ .
\end{split}
\end{equation}
This metric  describes the gravitational field for a wide range of compact objects, and only in the case of very dense ($ G\mathcal M \sim c^{2} \mathcal R$) or very rapidly rotating ($ G J \sim c^{3}\mathcal R ^ 2$) objects large discrepancies will appear.
\section{The Fock's approach}
\label{sec:fock}

Fock proposed in \cite{fock1959} a method to analyze Einstein's equations in the presence of matter and to derive approximate interior and exterior solutions. This approach takes into account the internal properties of the gravitational source, and reduces the problem of finding interior approximate solutions to the computation of some integrals that depend explicitly on the physical characteristics of the object. In this section, we present the main results of this approach, derive a particular interior approximate solution, and investigate the possibility of matching it with an exterior counterpart.

\subsection{The interior solution}

Fock's first-order approximation metric was recently re-derived and investigated by Abdildin \cite{1988mtge.book.....A} in a simple manner. Initially, this metric was written in its original form in a harmonic coordinate system \cite{dedonder1921, lanczos1923} as follows (a derivation of this metric is presented in the Appendix)
\begin{equation}\label{f1}
\begin{split}
ds^{2}=\left[c^{2}-2U+\frac{2U^{2}}{c^{2}}-\frac{2G}{c^{2}}\int\frac{\rho\left(\frac{3}{2}v^{2}+\Pi-U\right)+3p}{|\vec{r}-\vec{r'}|}\left(dx'\right)^{3}\right]dt^{2}\qquad \qquad \\ \qquad -\left(1+\frac{2U}{c^{2}}\right)\left({dx_1}^2+{dx_2}^2+{dx_3}^2\right)+\frac{8}{c^{2}}\left(U_1
dx_1+U_2 dx_2+U_3 dx_3\right)dt,
\end{split}
\end{equation}
where $U$ is the  Newtonian gravitational  potential, $\rho$  is the mass density of the body, $v$ is the speed of the particles inside the body, $\Pi$ is the elastic energy per unit mass, $p$ is the pressure, $\vec{U}$ is the gravitational vector potential. Notice that the quantities $\rho$, $v$, $\Pi$ and $U$  that characterize the inner structure of the source depend only on the ''inner" coordinates $x_i'$, which are defined inside the body only. To simplify the notations we omit the arguments that define this coordinate dependence.
The corresponding energy-momentum tensor is given as
\bea
T^{00} = \frac{\rho}{c^2}\left[ 1 + \frac{1}{c^2} \left(\frac{v^2}{2} + \Pi - U\right)\right] \ ,\quad T^{0i} = \frac{\rho}{c^2} v^i \ ,\quad T^{ij} =\frac{1}{c^2}\left( \rho v^iv^j + p\delta^{ij} \right) \ ,
\eea
where $\delta^{ij}$ is the Kronecker delta and $i,j=1,2,3$. Newton's potential satisfies the equation $ \nabla^2 U=-4\pi G\rho$. The solution of this equation that satisfies the asymptotically flatness condition at infinity can be written in the form of a volume integral:
\begin{equation}\label{}
U=G\int\frac{\rho}{|\vec{r}-\vec{r'}|}dx_1'dx_2'dx_3' \ .
\end{equation}
Furthermore, the vector potential must satisfy the equation $\nabla^2 U_{i}=-4\pi G\rho v_{i}$ whose general asymptotically flat solution can be represented as
\begin{equation}\label{}
U_{i}=G\int\frac{\rho v_{i}}{|\vec{r}-\vec{r'}|}dx_1'dx_2'dx_3' \ .
\end{equation}
Additional details about this metric can be found in \cite{2009GrCo...15....1A} and \cite{2009GrCo...15..141A}.

It is worth noticing that Chandrasekhar, using the Fock method, obtained in \cite{1965ApJ...142.1488C} a solution similar to (\ref{f1}) that later on was used by Hartle and Sharp in \cite{HS1967}. However, it is not difficult to show that Chandrasekhar's solution  is equivalent to (\ref{f1}). Indeed, the identification of the density
\begin{equation}\label{}
\rho=\rho_{Fock}=\rho_{Chandra}\left[1+\frac{1}{c^2}\left(3U+\frac{v^2}{2}\right)\right]
\end{equation}
at the level of the energy-momentum tensor allows one to calculate the corresponding metric functions that show the equivalence of the metrics.
Moreover, it has been shown in \cite{1965ApJ...142.1488C} that the solution for the non-rotating case can be matched with the well-known Schwarzschild solution, appropriately specialized to the case of spherical symmetry and hydrostatic equilibrium in the post Newtonian approximation.

\subsection{The exterior Fock metric for slow rotation and spherically symmetric distribution of mass}
In order to completely determine the metric, it is necessary to calculate the above integrals. Clearly, the result will depend on the internal structure of the body which is determined by the density $\rho$, pressure $p$ and velocity $v_i$ distributions. Once these functions are given, the calculation of the integrals can be performed in accordance with the detailed formalism developed by  Fock \cite{fock1959} and then extended and continued by Abdildin \cite{1985pft..conf...20A, abdildin2006} and Brumberg \cite{1991ercm.book.....B}. Consider, for instance, the case of a rotating sphere with total mass $M$. Then, the corresponding exterior metric in spherical-like (non harmonic) coordinates can be written as \cite{1985pft..conf...20A}
\be
\begin{split}
ds^{2}=\left[c^{2}-\frac{2GM}{r}-\kappa \frac{G {S_{0}^{2}}}{c^{2}M r^{3}}\left(1-3\cos^{2}\theta\right)\right]dt^{2}-\left(1+\frac{2GM}{c^{2}r }\right){dr}^2\\
-r^{2}\left({d\theta}^2+\sin^{2}\theta{d\phi}^2\right)+\frac{4G {S_{0}}}{c^{2}r}\sin^{2}\theta{d\phi}dt \ ,
\end{split}
\label{fockint}
\ee
where $S_{0}$ is the angular momentum of the body\footnote{Notice the typos in the sign in front of $S_{0}^{2}$ in Eqs. (1.78) and (1.79) of \cite{abdildin2006}}.
Here we added the constant $\kappa$ and verified that in fact the above metric is an approximate solution for any arbitrary real value of $\kappa$. This simple observation allows us to interpret Fock's procedure as a method to find out how the internal structure of the object influences the values of the external parameters. For instance, the total mass in the above metric is $M$ but it can decomposed as
\begin{equation}\label{f4}
M=m+\frac{\zeta}{c^2},
\end{equation}
where $m$ is the static mass of the body (for details see \cite{fock1959, 1991ercm.book.....B, HS1967}), and $\zeta$ is an arbitrary real constant which, as the constant $\kappa$, depends on the internal properties of the body. In particular, the cases of a liquid and a solid sphere have been analyzed in detail in \cite{1985pft..conf...20A, abdildin2006, 2009GrCo...15..141A} with the result
\begin{equation}\label{f5}
\zeta=\begin{cases}
\frac{8}{3}K+\frac{2}{3}\varepsilon,& \text{for a liquid sphere,}\\
4K+\frac{2}{3}\varepsilon,& \text{for a solid sphere,}
\end{cases}
\qquad\qquad
\kappa=\begin{cases}
\frac{4}{7},&\text{for a liquid sphere,}\\
\frac{15}{28},&\text{for a solid sphere.}
\end{cases}
\end{equation}
where $K$ is the rotational kinetic energy of the body and
$\varepsilon$ is the energy of the mutual gravitational attraction of
the particles inside the body. In Sec. \ref{sec:extfock}, we will briefly explain how to obtain the above values.

\subsection{The Kerr metric}

To describe the gravitational field of the rotating sphere outside the source, it seems physically reasonable to assume that the exterior vacuum metric be asymptotically flat. In this case, the first obvious candidate is the Kerr solution in the corresponding limit. The Kerr metric \cite{Kerr1963PhRvL..11..237K} in Boyer-Lindquist coordinates \cite{1967JMP.....8..265B, 1991ercm.book.....B} can be written as
\begin{equation}\label{k1}
\begin{split}
ds^{2}=\left(1-\frac{2\mu\varrho}{\varrho^{2}+a^{2}\cos^{2}\vartheta}\right)c^{2}dt^{2}-\frac{\varrho^{2}+a^{2}\cos^{2}\vartheta}{\varrho^{2}-2\mu\varrho+a^{2}}d\varrho^{2}-\left(\varrho^{2}+a^{2}{\cos^{2}\vartheta}\right)d\vartheta^{2}
\\-\left(\varrho^{2}+a^{2}+\frac{2\mu\varrho{a}^{2}\sin^{2}\vartheta}{\varrho^{2}+a^{2}\cos^{2}\vartheta}\right)\sin^{2}\vartheta{d\phi^{2}}-\frac{4\mu\varrho{a}\sin^{2}\vartheta}{\varrho^{2}+a^{2}\cos^{2}\vartheta}cdtd\phi\ ,
\end{split}
\end{equation}
where
\begin{equation}\label{k2}
\mu=\frac{GM}{c^{2}},\ a=-\frac{S_{0}}{Mc}.
\end{equation}
Expanding this metric to the order $\frac{1}{c^{2}}$, one obtains
\begin{equation}\label{k3}
\begin{split}
ds^{2}=\left[c^{2}-\frac{2GM}{\varrho}+\frac{2GM a^{2}}{\varrho^{3}}\cos^{2}\vartheta\right]dt^{2}-\left(1+\frac{2GM}{\varrho c^{2}}-\frac{a^{2}}{\varrho^{2}}\sin^{2}\vartheta\right)d\varrho^{2}
\\-\varrho^{2}\left(1+\frac{a^{2}}{\varrho^{2}}\cos^{2}\vartheta\right)d\vartheta^{2}-\varrho^{2}\left(1+\frac{a^{2}}{\varrho^{2}}\right)\sin^{2}\vartheta{d\phi^{2}}-\frac{4GMa}{\varrho c}\sin^{2}\vartheta{d\phi}{dt} \ .
\end{split}
\end{equation}
Furthermore, if we introduce new coordinates
$\varrho=\varrho(r,\theta)$, $\vartheta=\vartheta(r,\theta)$ by means of the equations
\be
\varrho=r-\frac{a^{2}\sin^{2}\theta}{2r},\ \ \vartheta=\theta-\frac{a^{2}\sin\theta\cos\theta}{2r^{2}}\ ,
\label{k5}
\ee
then the  approximate Kerr metric (\ref{k3}) can be reduced to the following form
\begin{equation}\label{k6}
\begin{split}
ds^{2}=\left[c^{2}-\frac{2GM}{r}-\frac{G {S_{0}^{2}}}{c^{2}M r^{3}}\left(1-3\cos^{2}\theta\right)\right]dt^{2}
-\left(1+\frac{2GM}{c^{2}r}\right){dr}^2\\-r^{2}\left({d\theta}^2+\sin^{2}\theta{d\phi}^2\right)+\frac{4GS_{0}}{c^{2}r}\sin^{2}\theta{d\phi}dt \ ,
\end{split}
\end{equation}
which coincides with the metric (\ref{fockint}) with $\kappa =1$. Consequently, the extended Fock metric (\ref{fockint}) can be interpreted as describing the exterior field of a rotating body to second order in the angular velocity. The advantage of using Fock's method to derive this approximate solution is that it allows the determination of the arbitrary constant $\kappa$. In fact, whereas $\kappa=\kappa_L= 4/7$ for a liquid sphere and $\kappa=\kappa_S= 15/28$ for a solid sphere, the value for the Kerr metric $\kappa=\kappa_K=1$ does not seem to correspond to a concrete internal model. On the other hand, all the attempts to find a physically meaningful interior Kerr solution have been unsuccessful. Perhaps the relationship with Fock's formalism we have established here could shed some light into the structure of the interior counterpart of the Kerr metric.

Furthermore, the coordinate transformation \cite{HT1968}
\be \varrho= R -\frac{a^2}{2R}\left[\left(1+\frac{2G \mathcal M}{c^2 R}\right)\left(1-\frac{G\mathcal M}{c^2 R}\right) - \cos^2\Theta \left(1 - \frac{2G \mathcal M}{c^2 R}\right)\left(1+\frac{3G \mathcal M}{c^2 R}\right)\right]\ ,
\ee
\be
\vartheta = \Theta - \frac{a^2}{2R^2}\left(1 + \frac{2G\mathcal M}{c^2 R}\right) \cos\Theta \sin\Theta \ ,
\ee
transforms the Kerr solution (\ref{k1}), expanded to second order in the angular momentum, (here one should set $G=c=1$) into the HT solution (\ref{ht1}) with $ J=-{\mathcal M}a $, $M=\mathcal M$ and a particular quadrupole parameter $Q=J^2/{\mathcal M}$.

In this way, we have shown that the extended Fock metric coincides for $\kappa=1$ with the approximate Kerr solution which, in turn, is equivalent to the exterior HT solution with a particular value of the quadrupole parameter. The fact that in the Kerr solution the quadrupole moment is completely specified by the angular momentum is an indication that it can be applied only to describe the gravitational field of a particular class of compact objects. A physically meaningful generalization of the Kerr solution should include a set of arbitrary multipole moments which are not completely determined by the angular momentum. In the next section, we present a particular exact solution characterized by an arbitrary quadrupole moment.


\section{A solution with quadrupole moment}
\label{sec:qua}

In this section, we will consider the case of deformed objects as, for example, a rotating ellipsoid. It is obvious that if the form of the body slightly deviates from spherical symmetry, it acquires multipole moments, in particular, a quadrupole moment; the moments of higher order are negligible, especially, for a slowly rotating ellipsoid. We will generalize Fock's metric so that the quadrupole moment appears explicitly from the integration of (\ref{f1}) and in the Newtonian potential. It should be mentioned that finding external and internal Newtonian potentials for a rotating ellipsoid is one of the classic problems of both physics and mathematical physics. Some examples for a homogeneous ellipsoid are considered in \cite{landau1962}, but the most comprehensive details on this matter are given in \cite{1969efe..book.....C} and more recently in \cite{2008rfe..book.....M}.  As for the exterior counterpart, there are several exact solutions \cite{1990CQGra...7..779C, 1990CQGra...7L.209M, 1992CQGra...9.2477M,2000PhRvD..61h1501M, 2006PhRvD..73j4038P} with quadrupole moment and rotation parameter  that could be used as possible candidates to be matched with the interior approximate solution. In this work, we limit ourselves to the study of a particular solution first proposed in \cite{1985PhLA..109...13Q} and then generalized in \cite{1986PhRvD..33..324Q,1991PhRvD..43.3902Q}.

\subsection{The exterior Fock solution}
\label{sec:extfock}

Let us consider the first-order approximation metric (\ref{f1}). It is convenient to use the notation $x_1'=x$, $x_2'=y$, and $x_3'=z$. In general, the fact that the mass density $\rho=\rho(x,y,z)$ is a function of the coordinates does not allow us to find explicit expression for the internal Newtonian potential. It is possible only by numerical integration. However, for the case of uniform density there is in the literature a reasonable number of exact solutions for rotating ellipsoids. Since we consider slow rotation and the weak field approximation, we can use the expansion for the Newtonian potential \cite{landau1962}, \cite{1999A&A...349..887L}
\begin{equation}\label{c2}
U(r, \theta)=G\int\frac{\rho}{|\vec{r}-\vec{r'}|}dx dy dz
=\frac{Gm}{r}+\frac{GD}{2r^{3}}P_{2}(\cos\theta) \ ,
\end{equation}
where $m$ is the rest mass of the ellipsoid, $D$ is the Newtonian quadrupole moment, $\theta$ is the angle between $r'=\sqrt{x^2+y^2+z^2}$
and $z$ --- axis. The first term in the expression above is the potential of a sphere and the second one is responsible for the deviation from spherical symmetry. If one takes the $z$ axis  as a rotating axis then the quadrupole moment is defined by
\begin{equation}\label{c3}
D=\int\rho(2z^{2}-x^{2}-y^{2})dxdydz.
\end{equation}
For the rotating ellipsoid with uniform density the quadrupole moment is well-known $D=2m\left(r_p^2-r_e^2\right)/5$, where $r_p$ and $r_e$ are the polar and equatorial radii of the ellipsoid, respectively. The mass of the ellipsoid is defined as the integral $m=\int\rho dxdydz$ that in the case of an ellipsoid with uniform density yields $ m=4\pi\rho r_e^{2}r_p/3$.
Note that the integration is carried out in the ranges of $0\leq x,y\leq r_e$ and $0\leq z \leq r_p$. Using the same procedure one may write the integral in Fock's metric as follows
\begin{equation}\label{c6}
\int\frac{\rho \left(\frac{3}{2}v^{2}+\Pi-U\right)+3p}{|\vec{r}-\vec{r'}|}  dx dy dz =\frac{\zeta}{r}+\frac{\mathcal
D}{2r^{3}}P_{2}(\cos\theta),
\end{equation}
where
\begin{eqnarray}\label{c7}
\zeta &=& \int\left[\rho\left(\frac{3}{2}v^{2}+\Pi-U\right)+3p\right]dxdydz\ , \\
\mathcal D &=& \int\left[\rho\left(\frac{3}{2}v^{2}+\Pi-U\right)+3p\right](2z^{2}-x^{2}-y^{2})dxdydz \ .
\end{eqnarray}
The quantity $\mathcal D/c^2$ is the relativistic correction to the Newtonian quadrupole moment $D$, i. e., the quadrupole moment due to rotation.
To evaluate the integrals we use the relation for a compressible elastic medium \cite{fock1959}
\begin{equation}\label{c10}
\rho\Pi-\rho U+p=\rho W\ ,
\end{equation}
where $W$ is the potential of the centrifugal forces determined by
\begin{equation}\label{c11}
W=\frac{(x^2+y^2)}{2}\Omega^2\ ,
\end{equation}
for rigid rotation  the angular velocity of the body $\vec{\Omega}=\{0,0,\Omega\}$ has only one component along $z$ axis and  $v^{2}=2W$.
Taking into account these expressions, the above shown equations reduce to the simple form
\begin{eqnarray}\label{c16}
\zeta&=&2\int\left[2\rho W+p\right]dxdydz \ , \\
\mathcal D&=&2\int\left[2\rho W+p\right](2z^{2}-x^{2}-y^{2})dx dy dz \ .
\end{eqnarray}

Furthermore, to evaluate these integrals we consider the following two cases that determine the inner structure of the body:

1) A liquid body with following the equation of internal motion \cite{fock1959}
\begin{equation}\label{c18}
\rho\frac{\partial }{\partial x_{i}}\left(U+W \right)=\frac{\partial p}{\partial x_{i}} \ .
\end{equation}

2) An absolute solid body with the following equation of internal motion \cite{1991ercm.book.....B}
\begin{equation}\label{c19}
\rho\frac{\partial U}{\partial x_{i}}=\frac{\partial p}{\partial x_{i}}\ .
\end{equation}
These are the equations of hydrostatic equilibrium which are adopted by Fock \cite{fock1959} and Brumberg \cite{1991ercm.book.....B} to describe the internal structure of the object. We limit ourselves to consider those cases in which the body rotates as a whole, in the manner of a rigid body. Then, for both
liquid and solid bodies the rotational kinetic energy takes the form
\begin{equation}\label{c20}
K=\int\rho Wdxdydz=\frac{I_{zz}\Omega^{2}}{2}\ ,
\end{equation}
where $I_{zz}$ is the moment of  inertia of the ellipsoid, which for a uniform density distribution,  is equal to
\begin{equation}\label{c20b}
I_{zz}=\int\rho (x^2+y^2)\,dx dy dz =\frac{2}{5}mr_e^{2}\ .
\end{equation}
The pressure can be expressed as
\begin{equation}\label{c21}
\int p\,dx dy dz =\begin{cases}
\frac{1}{3}\left(\varepsilon-2K\right),& \text{for a liquid body,}\\
\frac{1}{3}\varepsilon,& \text{for a solid body,}
\end{cases}
\end{equation}
where
\begin{equation}\label{c23}
\varepsilon=\frac{1}{2}\int\rho U\, dx dy dz \ ,
\end{equation}
represents the negative of the energy of mutual attraction of the constituent particles of the body. For a uniform density it has form
\begin{equation}\label{c23b}
\varepsilon=\frac{3Gm^2}{5\sqrt{r_e^{2}-r_p^{2}}}\arccos\frac{r_p}{r_e}.
\end{equation}
The second moments
\begin{equation}\label{c24}
K_{ik}=\int\rho W x_{i}'x_{k}' \, dx dy dz \ ,
\end{equation}
can be computed by using the above expressions. Then, for the second moments of the pressure we obtain (see \cite{1988mtge.book.....A})
\begin{equation}\label{c25}
\int p x_{i}'x_{k}'\, dx dy dz =\begin{cases}
\frac{1}{2}\chi_{ik}-\frac{2}{5}K_{ik},& \text{for a liquid body,}\\
-\frac{1}{2}K_{ik},& \text{for a solid body,}
\end{cases}
\end{equation}
where (more details can be found in \cite{1988mtge.book.....A})
\begin{equation}\label{c26}
\chi_{ik}=-\frac{2}{5}\int \rho x_{i}'x_{k}'x_{j}'\frac{\partial U}{\partial x_{j}'} \, dx dy dz \ .
\end{equation}

After calculating all the integrals we have
\begin{equation}\label{c28}
\zeta=\begin{cases}
\frac{8}{3}K+\frac{2}{3}\varepsilon,& \text{for a liquid body,}\\
4K+\frac{2}{3}\varepsilon,& \text{for a solid body,}
\end{cases}
\end{equation}
\begin{equation}\label{c29}
\mathcal D=\begin{cases}
\frac{28}{5}\frac{\kappa_{L}S_{0}^{2}}{I_{zz}^2}\left[\int \rho(x^2+y^2)(z^2-x^2)dxdydz \right]-\frac{4}{5}\int\rho(z^2-x^2)x_{j}'\frac{\partial U}{\partial x_{j}'}dxdydz ,&\text{for a liquid body,}\\
\frac{28}{5}\frac{\kappa_{S}S_{0}^{2}}{I_{zz}^2}\left[\int \rho(x^2+y^2)(z^2-x^2)dxdydz \right],&\text{for a solid
body,}
\end{cases}
\end{equation}
where $S_{0}$ is the angular momentum of the body, which is found from
\begin{equation}\label{c30}
\vec{S_{0}}=I_{zz}\vec{\Omega}\ ,
\end{equation}
and the numerical factors are
\begin{equation}\label{c31}
\kappa=\begin{cases}
\kappa_{L}=\frac{4}{7},&\text{for a liquid body,}\\
\kappa_{S}=\frac{15}{28},&\text{for a solid body.}
\end{cases}
\end{equation}
Unlike the Newtonian scalar potential, the vector potential can be easily calculated from
\begin{equation}\label{c32}
\vec{U}=\frac{G}{2r^{3}}\left[\vec{S_{0}}\times\vec{r}\right].
\end{equation}
Introducing the effective (total) mass as
\begin{equation}\label{c33}
M=m+\frac{\zeta}{c^2},
\end{equation}
for the Fock metric we obtain the following expression
\begin{equation}\label{34}
\begin{split}
ds^{2}=\left[c^{2}-2\left(\frac{GM}{r}+\frac{GD}{2r^{3}}P_2(\cos\theta)\right)+\frac{2}{c^{2}}
\left(\frac{GM}{r}+\frac{GD}{2r^{3}}P_2(\cos\theta)\right)^{2}-\frac{G\mathcal
D}
{c^{2}r^{3}}P_2(\cos\theta)\right]dt^{2}\\
-\left[1+\frac{2GM}{c^{2}r
}+\frac{GD}{c^{2}r^{3}}P_2(\cos\theta)\right]\left[{dr}^2+r^{2}({d\theta}^2+\sin^{2}\theta{d\phi}^2)\right]
+\frac{4G {S_{0}}}{c^{2}r}\sin^{2}\theta{d\phi}dt \ ,
\end{split}
\end{equation}
in harmonic coordinates. In order to write it in Schwarzschild like (standard) spherical coordinates one should use the coordinate transformation
\begin{equation}\label{c35}
r\rightarrow
R-\frac{GM}{c^{2}}, \quad \theta\rightarrow\Theta.
\end{equation}
which transforms the metric (\ref{34}) into
\begin{equation}\label{c36}
\begin{split}
ds^{2}=\left[c^{2}-\frac{2GM}{R}-\left(D+\frac{\mathcal
D}{c^{2}}\right)\frac{G}{R^{3}}P_2(\cos\Theta)-\frac{G^{2}DM}{c^{2}R^{4}}P_2(\cos\Theta)\right]dt^{2} +\frac{4G {S_{0}}}{c^{2}R}\sin^{2}\Theta{d\phi}dt\\
-\left[1+\frac{2GM}{c^{2}R}+\frac{GD}{c^{2}R^{3}}P_2(\cos\Theta)\right]{dR}^2
-\left[1+\frac{GD}{c^{2}R^{3}}P_2(\cos\Theta)\right]R^{2}({d\Theta}^2+\sin^{2}\Theta{d\phi}^2)
 \ ,
\end{split}
\end{equation}
where we have neglected quadratic terms in the quadrupole parameter $D$.
In the limiting case with vanishing rotation $S_0=0$ and vanishing quadrupole moment $D={\mathcal D}=0$, this metric represents the approximate Schwarzschild solution.

An examination of the metric (\ref{34}) shows that the rough approximation with $r_e\approx r_p\approx r_{sphere}$ and $S_{0}\neq0$ leads to the approximate Fock metric considered in Sec. \ref{sec:fock} with the total mass $M$,  for a slowly rotating spherically symmetric body with
\begin{equation}\label{c41}
D=0,\quad \mathcal D=-\frac{2\kappa S_{0}^{2}}{M}\ .
\end{equation}

It should be noted that an analogous result was obtained by Laarakkers and Poisson \cite{1999ApJ...512..282L}. They numerically computed the scalar quadrupole moment $\mathcal Q$ of rotating neutron stars for several equations of state (EoS). They found that for fixed gravitational mass $M$, the quadrupole moment is given as a simple quadratic fit
\begin{equation} \label{lp}
\mathcal Q=-\varkappa\frac{J^2}{Mc^2}
\end{equation}
where $J$ is the angular momentum of the star and $\varkappa$ is a dimensionless quantity that depends on the EoS.  Note that the scalar quadrupole moment $\mathcal Q$ of Laarakkers and Poisson is related to the one of Hartle and Thorne as follows $\mathcal Q=-Q$. The above quadratic fit reproduces $\mathcal Q$ with remarkable accuracy. The quantity $\varkappa$ varies between $\varkappa \approx 2$ for very soft EoS's and $\varkappa \approx 7.4$ for very stiff EoS's, for $M=1.4M_{Sun}$ as in neutron stars. This is considerably different from a Kerr black hole, for which $\varkappa=\kappa=1$ (see \cite{stergioulas, 1980RvMP...52..299T}).
Recently, the results of \cite{1999ApJ...512..282L} were modified taking into account the correct definition of multipole moments \cite{2012PhRvL.108w1104P}. Therefore the value of the $\varkappa$ parameter in the numerical fit (\ref{lp}) is slightly different from that given in \cite{1999ApJ...512..282L}.
In our case, we have  similar, but not the same results, since the Fock solution is not valid in the limit of strong gravitational fields (like in neutron stars) and fast rotation. The values for the constant $\kappa$ are obtained from qualitative analyses in the limit of weak field and slow rotation. In order to find exact values for $\kappa$ one should specify the EoS's and perform numerical integrations. This task, however, is out of the scope of the present work.

\subsection{The exterior Quevedo-Mashhoon solution}
\label{sec:qm}

In this section, we study the general metric describing the gravitational field of a rotating deformed mass found by Quevedo and Mashhoon \cite{1985PhLA..109...13Q,1986PhRvD..33..324Q,1989PhRvD..39.2904Q, 1991PhRvD..43.3902Q}, which is a stationary axisymmetric solution of the vacuum Einstein's equations belonging to the class of Weyl-Lewis-Papapetrou \cite{1917AnP...359..117W, 1932RSPSA.136..176L, Papapetrou1966}. For the sake of simplicity we consider here a particular solution involving only four parameters: the mass parameter $M$, the angular momentum parameter $a$, the quadrupole parameter $q$, and the  additional Zipoy-Voorhees \cite{1966JMP.....7.1137Z, 1970PhRvD...2.2119V} constant $\delta$. For brevity, in this section we use geometric units with $G=c=1$. The corresponding line element in spheroidal coordinates $(t, r,\theta,\phi)$ with $r\geq \sigma+M_0,\ 0\leq \theta \leq \frac{\pi}{2}$ is given by \cite{1985PhLA..109...13Q}
\begin{eqnarray}\label{q1}
ds^2=f(dt-\omega d\phi)^2- \frac{1}{f}
\bigg\{e^{2\gamma}\left(d\theta^2+\frac{dr^2}{r^2-2M_0r+a^2}\right)\left[(M_0-r)^2-(M_{0}^2-a^2)\cos^2\theta\right] \nonumber \\
+\left(r^2-2M_0r+a^2\right)\sin^{2}\theta d\phi^2\bigg\},\quad
\end{eqnarray}
where $f,\ \omega$ and $\gamma$ are functions of $r$ and $\theta$ only, and $\sigma$ is a constant. They have the form $[x=(r-M_0)/\sigma,\ y=\cos\theta]$
\begin{eqnarray}\label{q2}
f=\frac{\tilde{R}}{L}e^{-2q\delta P_{2}Q_{2}},\qquad \omega=-2a-2\sigma\frac{\mathfrak{M}}{\tilde{R}}e^{2q\delta P_{2}Q_{2}},\qquad e^{2\gamma}&=&\frac{1}{4}\left(1+\frac{M}{\sigma}\right)^{2}\frac{\tilde{R}}{(x^{2}-1)^{\delta}}e^{2\delta^{2}\hat{\gamma}},
\end{eqnarray}
where
\begin{eqnarray}\label{q5}
\tilde{R}=a_{+}a_{-}+b_{+}b_{-},\quad\quad\; L=a_{+}^{2}+b_{+}^{2},\quad\quad\quad\quad\quad\quad\quad\quad\quad\quad\quad\quad\quad\quad\quad\\
\mathfrak{M}=(x+1)^{\delta-1}\left[x(1-y^{2})(\lambda+\eta)a_{+}+y(x^{2}-1)(1-\lambda\eta)b_{+}\right],\quad\quad\quad\quad\quad\\
\hat{\gamma}=\frac{1}{2}(1+q)^{2}\ln\frac{x^{2}-1}{x^{2}-y^{2}}+2q(1-P_{2})Q_{1}+q^{2}(1-P_{2})[(1+P_{2})(Q_{1}^{2}-Q_{2}^{2})\\
+\frac{1}{2}(x^{2}-1)(2Q_{2}^{2}-3xQ_{1}Q_{2}+3Q_{0}Q_{2}-Q_{2}^{\prime})].\nonumber
\end{eqnarray}
Here $P_{l}(y)$ and $Q_{l}(x)$ are Legendre polynomials of the first and second kind, respectively. Furthermore
\begin{eqnarray}\label{q8}
a_{\pm}&=&(x\pm1)^{\delta-1}[x(1-\lambda\eta)\pm(1+\lambda\eta)],\\
b_{\pm}&=&(x\pm1)^{\delta-1}[y(\lambda+\eta)\mp(\lambda-\eta)],
\end{eqnarray}
with
\begin{eqnarray}
\lambda&=&\alpha(x^{2}-1)^{1-\delta}(x+y)^{2\delta-2}e^{2q\delta\delta_{+}},\\
\eta&=&\alpha(x^{2}-1)^{1-\delta}(x-y)^{2\delta-2}e^{2q\delta\delta_{-}},\\
\delta_{\pm}&=&\frac{1}{2}\ln\frac{(x\pm y)^{2}}{x^{2}-1}+\frac{3}{2}(1-y^{2}\mp xy) +\displaystyle \frac{3}{4}[x(1-y^{2})\mp y(x^{2}-1)]\ln\frac{x-1}{x+1}\ .
\end{eqnarray}
Moreover, $\alpha$ and $\sigma$ are constants defined as
\begin{equation}\label{q11}
\alpha=\frac{\sigma-M}{a},\quad\quad\sigma=\sqrt{M^{2}-a^{2}}.
\end{equation}
The physical meaning of the parameters entering this metric can be investigated in an invariant manner by calculating the Geroch-Hansen \cite{1970JMP....11.2580G, 1974JMP....15...46H} moments:
\begin{eqnarray}
M_{2k+1}=J_{2k}=0,\quad k=0,1,2,\ldots,\quad\quad\quad\quad\quad\quad\quad\quad\quad\quad\quad\quad\quad\quad\quad\quad\quad\quad\quad\quad\quad\quad\quad\quad\\
M_{0}=M+\sigma (\delta-1),\quad J_{1}=Ma+2a\sigma(\delta-1),\quad\quad\quad\quad\quad\quad\quad\quad\quad\quad\quad\quad\quad\quad\quad\quad\quad\quad\quad\\
M_{2}=-Ma^{2}+\frac{2}{15}q\sigma^{3}-\frac{1}{15}\sigma(\delta-1)\left[45M^{2}+15M\sigma(\delta-1)-(30+2q+10\delta-5\delta^2)\sigma^{2}\right],\quad\\
J_{3}=-Ma^{3}+\frac{4}{15}aq\sigma^{3}-\frac{1}{15}a\sigma(\delta-1)\left[60M^{2}+45M\sigma(\delta-1)-2\sigma^{2}(15+2q+10\delta-5\delta^{2})\right].
\end{eqnarray}
The vanishing of the odd gravitoelectric $(M_{n})$ and even gravitomagnetic $(J_{n})$ multipole moments is a consequence of the reflection symmetry with respect to the equatorial plane $\theta=\pi/2$. Note that in the limiting case $\delta =1$, $M_{0}=M$ is the total mass of the body, $a$ represents the specific angular momentum, and $q$ is related to the deviation from spherical symmetry. All higher multipole moments can be shown to depend only on the parameters $M,\ a$, and $q$. In general, we see that the Zipoy-Voorhees parameter is related to the quadrupole moment of the source. In fact, even in the limiting static case with $a=0$ and $q=0$, the only non-vanishing parameters are $M=\sigma$ and $\delta$ so that all gravitomagnetic multipoles vanish and one obtains $M_0=M\delta$ and $M_2=-\frac{1}{3}M^{3}\delta(\delta^{2}-1)$ --- the quadrupole moment that indicates a deviation from spherical symmetry. Some geometrical properties of (\ref{q1}) versus particle motion and tidal indicators in this spacetime were explored in \cite{bini2009} and \cite{2012CQGra..29n5003B}, respectively.

Consider the limiting cases of the QM solution. For vanishing quadrupole parameter, $q=0$, $\delta=1$, and vanishing angular momentum $a=0, \ \alpha=0,$ and $\sigma=M$, one recovers the Schwarzschild solution with the following metric functions:
\begin{equation}\label{}
f=1-\frac{2M}{r}, \quad \omega=0, \quad \gamma=\frac{1}{2}\ln\frac{r(r-2M)}{(M-r)^{2}-M^{2}\cos^{2}\theta}.
\end{equation}
For vanishing quadrupole parameter and $\delta=1$, one recovers the Kerr solution (\ref{k1}) with $\vartheta\to\theta$ and $\varrho\to r$ and functions
\begin{equation}\label{q13}
f=1-\frac{2Mr}{r^{2}+a^{2}\cos^{2}\theta},
\quad \omega=\frac{2aMr\sin^{2}\theta}{r^{2}-2Mr+a^{2}\cos^{2}\theta},
\quad
\gamma=\frac{1}{2}\ln\frac{r(r-2M)+a^{2}\cos^{2}\theta}{(M-r)^{2}-(M^{2}-a^{2})\cos^{2}\theta}\ .
\end{equation}
The above limiting cases show that this solution describes the exact exterior field a rotating deformed object. To compute the case of a slowly rotating and slightly deformed body we choose the Zipoy-Voorhees parameter as $\delta=1+sq$, where $s$ is a real constant. Then, expanding the metric (\ref{q1}) to first order in the quadrupole parameter $q$ and to second order in the rotation parameter $a$, we obtain
\begin{multline}\label{qma1}
f=1-\frac{2M}{r}+\frac{2a^{2}M\cos^{2}\theta}{r^{3}}+q(1+s)\left(1-\frac{2M}{r}\right)\ln\left(1-\frac{2M}{r}\right)
\\+3q\left(\frac{r}{2M}-1\right)\left[\left(1-\frac{M}{r}\right)\left(3\cos^{2}\theta-1\right)+\left\{\left(\frac{r}{2M}-1\right)(3\cos^{2}\theta-1)-\frac{M}{r}\sin^{2}\theta\right\}\ln\left(1-\frac{2M}{r}\right)\right],
\end{multline}
\begin{equation}\label{qma2}
\omega=\frac{2aMr\sin^{2}\theta}{r-2M},
\end{equation}
\begin{multline}\label{qma3}
\gamma=\frac{1}{2}\ln\frac{r(r-2M)}{(r-M)^{2}-M^{2}\cos^{2}\theta}+\frac{a^{2}}{2}\left[\frac{M^{2}\cos^{2}\theta\sin^{2}\theta}{r(r-2M)((r-M)^{2}-M^{2}\cos^{2}\theta)}\right]
\\+q(1+s)\ln\frac{r(r-2M)}{(r-M)^{2}-M^{2}\cos^{2}\theta}-3q\left[1+\frac{1}{2}\left(\frac{r}{M}-1\right)\ln\left(1-\frac{2M}{r}\right)\right]\sin^{2}\theta.
\end{multline}

The further simplification $s=-1$, and the coordinate transformation
\cite{HT1968, bini2009, 1991NCimB.106..545M}
\begin{multline}\label{}
r=R+{\mathcal M}q+\frac{3}{2}{\mathcal M}q\sin^{2}\Theta\left[\frac{R}{\mathcal M}-1+\frac{R^{2}}{2{\mathcal M}^{2}}\left(1-\frac{2{\mathcal M}}{R}\right)\ln\left(1-\frac{2{\mathcal M}}{R}\right)\right]
\\-\frac{a^{2}}{2R}\left[\left( 1+\frac{2{\mathcal M}}{R}\right)\left( 1-\frac{{\mathcal M}}{R}\right)-\cos^{2}\Theta \left( 1-\frac{2{\mathcal M}}{R}\right)\left( 1+\frac{3{\mathcal M}}{R}\right)\right]
\end{multline}
\begin{equation}\label{}
\theta=\Theta-\sin\Theta\cos\Theta\left\{\frac{3}{2}q\left[2+\left(\frac{R}{{\mathcal M}}-1\right)\ln\left(1-\frac{2{\mathcal M}}{R}\right)\right]+\frac{a^{2}}{2R}\left(1+\frac{2{\mathcal M}}{R}\right)\right\}
\end{equation}
transforms the approximate QM solution (\ref{qma1})--(\ref{qma3}) into
\begin{equation}
\label{qmapprox}
\begin{split}
ds^2=\left[1-\frac{2G{M(1-q)}}{c^{2}R}+\frac{2G}{c^{2}R^{3}} \left(\frac{J^2}{M}-\frac{4}{5}qM^3\right) \left(1+\frac{GM(1-q)}{c^2R}\right)P_2(\cos\Theta)\right]
c^2dt^2 \\
-\frac{4GMa}{c^{2}R}\sin^2\Theta dt d\phi
-\left[1+\frac{2G{ M(1-q) }}{c^{2}R}-\frac{2G}{c^{2}R^{3}}  \left(\frac{J^2}{M}-\frac{4}{5}qM^3\right)  P_2(\cos\Theta)\right]dR^2 \\
-\left[1-\frac{2G}{c^{2}R^{3}}  \left(\frac{J^2}{M}-\frac{4}{5}qM^3\right) P_2(\cos\Theta)\right]R^2(d\Theta^2+\sin^2\Theta d\phi^2)\ .
\end{split}
\end{equation}
Here we introduced again all the necessary constants $G$ and $c$ in order to compare our results with previous metrics. Finally, if we redefine the parameters $M$, $a$, and $q$ as
\begin{equation}\label{mjq}
{\mathcal M}=M(1-q),\ \ J=-Ma,\ \ Q=\frac{J^{2}}{M}-\frac{4}{5}M^{3}q\ ,
\end{equation}
the approximate metric (\ref{qmapprox}) coincides with the exterior HT metric (\ref{ht4}) and, consequently, can be matched with the interior HT metric discussed in \ref{sec:ht}.

The above metric is equivalent to the exterior extended Fock metric discussed in the previous subsection. To see this one has to consider the exterior solution (\ref{c36}) which is written in the same coordinates as the exterior solutions (\ref{qmapprox}) and  (\ref{ht4}). It is convenient to show first the equivalence with the exterior HT solution (\ref{ht4}) that yields the conditions
\begin{equation}\label{c40}
\mathcal M=M, \qquad J=S_{0}, \qquad Q=-\frac{1}{2}\left(D+\frac{\mathcal D}{c^{2}}\right).
\end{equation}
The equivalence with the approximate QM solution (\ref{qmapprox}) follows then from the comparison of Eqs.(\ref{mjq}) and
(\ref{c40}). We obtain
\begin{equation}\label{c42}
q=\frac{5}{8}\frac{c^4}{G^2}\frac{1}{M^3}\left[D+\frac{1}{c^2}\left(\mathcal D+\frac{2 J^{2}}{M}\right)\right]\ ,
\end{equation}
and for vanishing $D$
\begin{equation}\label{c43}
q=\frac{5}{4}\frac{c^2}{G^2}\frac{J^{2}}{M^4}\left(1-\kappa\right).
\end{equation}
This result is in accordance with the limiting case of the Kerr metric for which we obtained that $\kappa= 1$ and hence $q=0$.

Thus we come to the conclusion that in the limit of a slowly rotating and slightly deformed body the QM approximate solution is equivalent to the exterior Fock solution.

\section{Conclusions}
\label{sec:con}

In this work, we studied the gravitational field of slowly rotating, slightly deformed astrophysical compact objects. We presented  the main exact and approximate solutions of Einstein's equations that can be used to describe the interior and the exterior gravitational field. In particular, we presented the method proposed by Hartle and Thorne to find interior and exterior approximate solutions, and the method proposed by Fock to derive approximate interior and exterior solutions. We derived an extension of the approximate exterior Fock metric that takes into account  up to the first order the contribution of a quadrupole parameter that describes the deviation of the body from spherical symmetry.  A particular parameter that enters the extended Fock metric turns out to have very specific values in the case of a liquid sphere and a solid sphere. In the case of the approximate Kerr metric, this parameter does not seem to correspond to any known interior model analyzed in the framework of Fock's formalism.

We found that a particular QM solution, which in general possesses an infinite set of gravitational and electromagnetic multipole moments, contains the exact Kerr metric and the approximate HT metric as special cases. Moreover, since the HT solution is endowed with its interior counterpart, we conclude that the approximate QM solution (to the second order in the angular momentum and to the first order in the quadrupole parameter) can be matched with the interior HT solution, indicating that it can be used to correctly describe the gravitational field of astrophysical compact objects. Moreover, we showed that the  explicit form of the exterior Fock metric is equivalent to the approximate exterior QM solution.

To avoid the technical problems that are usually found in the process of matching solutions \cite{2012mgm..conf...35Q}, we use the same set of coordinates inside and outside the body. In the cases presented here, this can be done in a relative easy way only because all the coordinate transformations are not calculated exactly, but with the same approximation as the metric functions. This approach allows us to reduce the matching problem to the comparison of the metrics on the matching surfaces in such a way that only algebraic conditions appear. Using this method, we could show that the approximate Kerr metric cannot be matched with an interior Fock solution. However, if we take into account an additional quadrupole parameter, the matching of the extended Fock metric can be carried out by using as exterior counterpart a particular approximate QM solution that contains the Kerr metric as special case. We conclude that the quadrupole parameter offers an additional degree of freedom that allows the matching. A first step in this direction was recently taken forward in \cite{2011GReGr..43.1141Q}.
It would be interesting to see if this is also true in the case of exact solutions. This could shed some light into the problem of finding a realistic gravitational source for the Kerr metric, a long-standing problem of classical general relativity.

\section*{Acknowledgements}
We would like to thank the ICRANet for support. One of us (K.B.) wishes to express his deep gratitude to M. Abishev for his guidance and encouragement during the course of this research, to I. Siutsou for  comments and remarks on an earlier version of this paper which led to improvements, and to the  ICRANet for financial support. This work was supported in part by DGAPA-UNAM, grant No. 106110.

\appendix

\section{Derivation of the Fock extended metric}\label{sec:appabd}

In this Appendix, we present a review of the derivation of a generalization of Fock's metric, based upon the approach formulated by Abdildin in \cite{abdildin2006}.
The original approximate metric derived by Fock in \cite{fock1959} can be written as
\begin{equation}\label{fock1}
ds^{2}=\left(c^{2}-2U\right)
dt^{2}
-\left(1+\frac{2U}{c^{2}}\right)\left({dx_1}^2+{dx_2}^2+{dx_3}^2\right)+\frac{8}{c^{2}}\left(U_1
dx_1+U_2 dx_2+U_3 dx_3\right)dt,
\end{equation}
where $U$ is the  Newtonian gravitational  potential that satisfies the equation $\nabla^2 U = - 4\pi G \rho$, where $\rho$ represents the matter density of the gravitational source. Moreover, the gravitational vector potential  $\vec{U}$ satisfies the equation  $\nabla^2 U_i = - 4\pi G \rho v_i$, where $v_i$ are the components of the 3-velocity of the particles inside the source. The coordinates $x^\mu$ are harmonic functions satisfying the D'Alambert equation $\Box x^\mu=0$.

As noticed by Abdildin, the metric (\ref{fock1}) presents certain difficulties. First, the components $g_{0i}$ and $g_{ij}$ contain a relativistic contribution that is absent in the component $g_{00}$. Second, if we use the metric (\ref{fock1}) to investigate the motion of test particles in a central potential, we obtain an expression for the perihelion shift that differs from the correct one by a factor of 1/2. Finally, in the case of a static field or for Gaussian-like coordinate systems $Udt^2 \sim dx_1^2 + dx_2^2 + dx_3^2$, i.e., the relativistic correction of $g_{00}$ must be of the same order as that of $g_{ij}$.

From the above observations it follows that it is necessary to consider a more appropriate expression for the component
\be
g^{00}= \frac{1}{c^2}+\frac{2U}{c^4} + \frac{\Phi}{c^6}\ ,
\ee
where $\Phi$ is an unknown function which must satisfy the corresponding approximate Einstein equation in harmonic coordinates
\be
R^{00}=\frac{1}{2}\nabla^2 g^{00} - \frac{2U}{c^6}\nabla^2 U - \frac{2}{c^6}\sum_i \left(\frac{\partial U}{\partial x_i}\right)^2 = - \frac{8\pi G}{c^2}
\left( T^{00} -\frac{1}{2}g^{00} T\right) \ .
\label{eins00}
\ee
As for the components of a energy-momentum tensor, in the case of an elastic source one can use the expressions
\be
T^{00}  =  \frac{\rho}{c^2}\left[ 1 + \frac{1}{c^2} \left(\frac{v^2}{2} + \Pi - U\right)\right] \ ,\quad
T^{0i}   =  \frac{\rho}{c^2} v^i \ ,\quad T^{ij} =\frac{1}{c^2}\left( \rho v^iv^j + p\delta^{ij} \right) \ ,
\label{emt}
\ee
where $\Pi$ is the elastic energy. It is then straightforward from Eqs.(\ref{eins00}) and (\ref{emt}) to conclude that
\be
\Phi = 2U^2 + 2 G \int \frac{\rho\left(\frac{3}{2}v^2 + \Pi - U\right) +3p  }{|\vec{r}-\vec{r}'|} (dx')^3.
\ee
Consequently, the generalized approximate metric is
\begin{equation}\label{abd1}
\begin{split}
ds^{2}=\left[c^{2}-2U+\frac{2U^{2}}{c^{2}}-\frac{2G}{c^{2}}\int\frac{\rho\left(\frac{3}{2}v^{2}+\Pi-U\right)+3p}{|\vec{r}-\vec{r'}|}\left(dx'\right)^{3}\right]dt^{2}\qquad\qquad
\\-\left(1+\frac{2U}{c^{2}}\right)\left({dx_1}^2+{dx_2}^2+{dx_3}^2\right)+\frac{8}{c^{2}}\left(U_1
dx_1+U_2 dx_2+U_3 dx_3\right)dt \ .
\end{split}
\end{equation}
This form of the metric overcomes all the difficulties mentioned above for the original Fock metric (\ref{fock1}), and is used everywhere in the present work to obtain the correct approximations.


%

\end{document}